# Demonstration of a Plasmonic Dimple Lens for Nanoscale Focusing of Light

Hyojune Lee[1], Shantha Vedantam[1], Japeck Tang[1], Josh Conway[1,2], Matteo Staffaroni[1], and Eli Yablonovitch[1,3]

[1]*Electrical Engineering Department, University of California, Los Angeles, California 90095*
[2]*The Aerospace Corporation, El Segundo, California 90245*
[3]*Department of Electrical Engineering & Computer Sciences, University of California, Berkeley, California 94720*

Focusing electromagnetic energy to sub-wavelength dimensions has become an increasingly active field of research for a variety of applications such as Heat Assisted Magnetic Recording (HAMR), nanolithography, and nanoscale optical characterization of biological cells and single molecules using near-field scanning optical microscopy (NSOM) technique. Double-sided surface plasmons in a metal-insulator-metal (MIM) geometry have been shown to have very small wavelengths for dielectric of thickness of less than 10 nm. A tapered dielectric structure sandwiched between metal, can be used to efficiently couple electromagnetic energy from free space photons to the plasmonic wavelengths at the nanoscale. In this paper, we present the fabrication and characterization of a novel MIM plasmonic lens structure.

## 1. Introduction

Conventional optics can only focus light to a diffraction-limited spot size of approximately $\lambda/2n$ in a medium with refractive index $n$. For visible wavelengths, the diffraction-limited spot is on the order of 100 nm even in a high index transparent material like diamond ($n\sim2.4$). In order to further focus light to smaller dimensions, one approach has been to use the confinement of electric fields that occurs around a sharp tip[1,2]. Near-field scanning optical microscopy (NSOM) is another standard technique for confining light to minimum spot size of ~100 nm using metalized pulled optical fibers with a nanoscale aperture. The transmission through these tapers is strictly evanescent and further reduction of the aperture size results in significant reduction in throughput[3,4]. An improvement in throughput can be obtained using C-shaped apertures, which have been shown to have better power transmission and smaller spot sizes than rectangular or square apertures of similar area[5].

Another approach to better confine electromagnetic energy has been to take advantage of the shorter wavelengths of optical waves running along a metal-dielectric interface, known as surface plasmons[6]. Grating couplers, similar to antennas, can been used to couple free space light into surface plasmons. Furthermore, a circular grating can be used to focus these surface plasmons toward a spot at the center of the circle. This is analogous to a lens for free space light, which results in a higher intensity at the focal point[7]. Unfortunately, single-sided plasmons have a fixed spot size that is determined by the wavelength of light and the dielectric constants of the metal and dielectric. In contrast, the wavelength of double-sided plasmons can be tuned geometrically by varying the thickness of the gap. It has been shown that the plasmon wavelength of the anti-symmetric modes in metal-insulator-metal structure decreases with the thickness of the dielectric. Thus, one can further reduce the spot size by employing double-sided plasmons in a structure with a thin dielectric. However, this introduces additional complication of requiring a way to efficiently couple energy from a large dielectric thickness to the smaller thickness, such as via a tapered structure[8,9]. The current work presents the basic theory, fabrication, and characterization of a 3-D tapered metal-insulator-metal structure for focusing double-sided plasmons to nanometer scale dimensions.

## 2. Theoretical Background

It has been shown[8,10] that double-sided surface plasmons can confine light to the nanometer scale by decreasing the dielectric thickness of the metal-insulator-metal stack. Figure 1 shows the dispersion relation of Au-SiO$_2$-Au stack for various dielectric thicknesses using the optical constants from reference 11. It is evident from the graph that for a fixed wavelength of operation, say 633 nm (1.96 eV), the double-sided surface plasmon wavelength can be varied from 390 nm to less than 20 nm by varying the thickness of SiO$_2$ from 100 nm down to 1 nm.

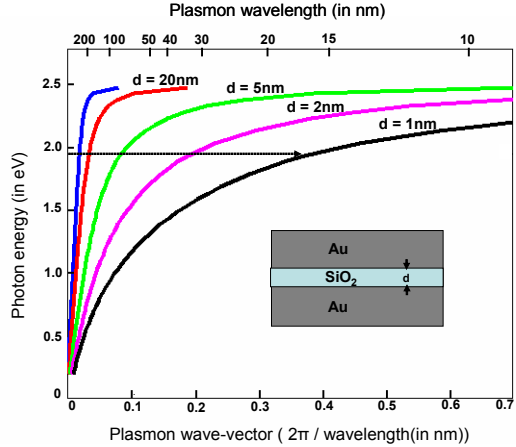

Figure 1 Calculated dispersion relations for the Au-SiO$_2$-Au stack for various thicknesses of the SiO$_2$ layer.

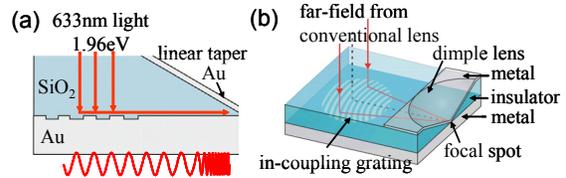

Figure 2 (a) Cross-sectional view schematic of the plasmonic dimple lens. Thickness of SiO$_2$ layer tapers in vertical direction. (b) Perspective view of the plasmonic dimple lens.

In this work, we employ a tapered MIM structure as shown in Figure 2a, to focus double-sided surface plasmons in the direction normal to the layers of the stack. An optimal taper angle ensures efficient coupling of electromagnetic energy from larger plasmonic wavelengths to nanoscale plasmonic wavelengths with minimum resistive and reflective energy losses[8]. A circular grating coupler is designed to couple free space visible light into single-sided surface plasmons before they traverse the taper to the out-coupling facet. As the plasmons propagate across the taper, their phase velocity is reduced in proportion to their wavelength. The simple foundations of Snell's Law still hold in this regime, making focusing in the direction normal to that of Figure 2a impossible. The macroscopic solution to this problem is the classical immersion lens, whose analogue in the plasmonic regime is formed by rotating the taper about its focus, as depicted in Figure 2b. We refer to this focusing structure as a plasmonic dimple lens. We note that this structure is fundamentally different from its macroscopic counterpart in that the effective index is changing continuously and that this change can be larger than an order of magnitude. These traits lead to unique properties such as self-focusing as stray light is guided to the extremely high index focus. The short plasmonic wavelengths at the end of the taper give rise to evanescent energy density at the focal spot on the out-coupling facet. Therefore, optical focusing is achieved in all three dimensions that confines energy to nanoscopic volume. The design parameters and field enhancement estimates can be found in reference 8.

## 3. Fabrication and Measurement

The critical dimension determining the spot size of the focused plasmonic energy structure shown in Figure 2 is the final thickness of the dielectric at the end of the taper. The fabrication of a three-dimensional circular taper in a dielectric terminating with a thickness of a few nanometers poses a significant challenge. For small dielectric thickness in a MIM structure, double-sided surface plasmons have short wavelengths (see Figure 1), which have large resistive energy losses and hence short propagation lengths[8]. Since double-sided plasmons are confined to the two metal-dielectric interfaces, they suffer additional energy loss by scattering via interface roughness. It has been observed that e-beam evaporated gold and silver surfaces typically have surface roughness on the order of 2-3 nm rms due to the formation of grains[12,13]. To keep this scattering loss low, it is important to ensure smooth surfaces at the two metal-dielectric interfaces that support plasmons. In view of these considerations, we fabricated the plasmonic dimple lens as Au-insulator-Au structure where the dielectric consisted of a layer of polymethylmethacrylate (PMMA) and a layer of thin silicon nitride. The maturity in silicon processing yields high quality amorphous oxide and nitride films on a silicon surface with a few tens of angstroms in thickness[14,15]. Also much progress has been made in tailoring smooth grayscale profiles in e-beam and photosensitive resists[16-20]. Our approach is to fabricate the taper profile in PMMA (a positive e-beam resist) on top of a thin layer of silicon nitride that determines the minimum dielectric thickness of the plasmonic dimple lens. The refractive index of the PMMA is 1.49-1.52 and is close to that of SiO$_2$, which has been used to determine the design parameters of the plasmonic lens. Our process sequence also ensures that gold is deposited onto smooth dielectric surfaces thereby minimizing the scattering energy loss owing to interface roughness.

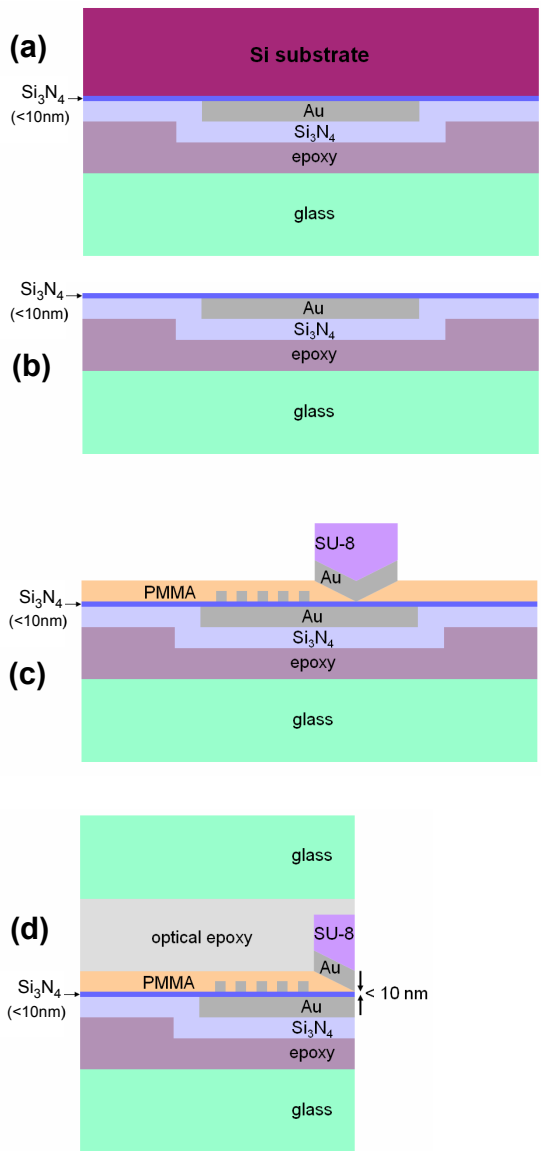

Figure 3 (a) Au islands patterned on 10 nm LPCVD SiN on a Si-substrate. These islands are coated with 300 nm of PECVD SiN and bonded to a glass using silica-filled epoxy. (b) Si-substrate is completely etched away with HNO$_3$+HF after thinning it down with DRIE process. (c) Semi-circular grating coupler and dimple profile in PMMA are patterned with e-beam lithography and second layer of Au is deposited and patterned. (d) The sample is bonded to glass using optically transparent epoxy and this stack is mechanically polished from one edge until the point when the circular dimple profile in PMMA is polished halfway through.

photolithography, e-beam evaporation and liftoff process (Figure 3a). These islands constitute the lower Au layer in the Au-dielectric-Au stack of Figure 2a. They also include the alignment markers for subsequent e-beam lithography steps, and the endpoint detection markers for the final edge-polishing step.

About 300 nm of silicon nitride was deposited on top of these islands by plasma enhanced chemical vapor deposition (PECVD) process to reduce stress on the thin LPCVD silicon nitride during the subsequent bulk silicon removal step. The patterned side of the sample was bonded to a glass piece using low stress silica-filled epoxy for support during the silicon substrate removal. This stack is shown in Figure 3a. The silicon substrate was thinned down to less than 50 μm using deep reactive ion etching (DRIE). The remaining silicon was removed by wet etching in a 1:1 mixture of HF and HNO$_3$ as depicted in Figure 3b. This mixture has a good selectivity between silicon and LPCVD silicon nitride.

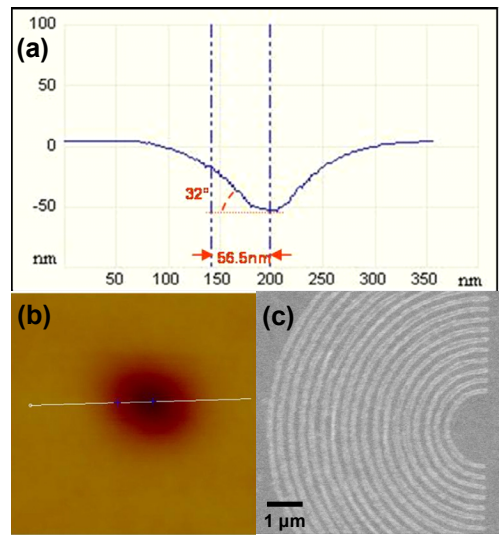

Figure 4 (a) Cross section and (b) 2D surface image acquired from the topographic AFM scan of the dimple profile in PMMA. The size of the scan is 400 nm x 400 nm (c) SEM image of the circular grating coupler.

Thin silicon nitride of ~10 nm thickness, was deposited on a standard silicon wafer, using low-pressure chemical vapor deposition (LPCVD) process. Gold islands were then patterned on this smooth silicon nitride surface using

A semi-circular Au grating of 30 nm height and 390 nm period was patterned using e-beam lithography, e-beam evaporation, and liftoff process. A circular dimple profile was then formed in a 100 nm layer of PMMA with a single spot exposure using e-beam lithography. Both the grating and the dimple were inspected using a scanning electron microscope (SEM) and an atomic force microscope (AFM), respectively, as can be seen in Figure 4. To form the top Au layer of Au-dielectric-Au stack as shown in Figure 2a, 100 nm layer of Au was evaporated on top of the PMMA layer using e-beam evaporation. This Au layer was then etched to gain optical access to the grating using Ar-ion sputter etch with e-beam patterned SU-8 (a negative e-beam resist) as a mask. Figure 3c depicts the stack at this stage. The structure is then bonded to another piece of glass using optically transparent epoxy for additional support during the mechanical polishing step. Finally, the sample is polished from the edge until the center of the dimple is exposed at the polished edge facet (Figure 3d). The topographic and phase-shift images from the AFM scan of the polished edge facet are shown in Figure 5.

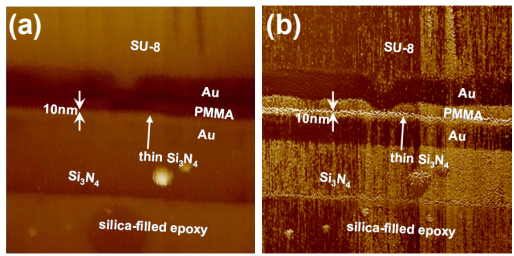

Figure 5 (a) Topographic and (b) phase-shift AFM image of the out-coupling edge after mechanical edge-polishing. The size of the scan is 1.5 μm x 1.5 μm. Good polishing results with minimal polishing relief make distinguishing different layers in the topographic image difficult. Phase-shift image makes identification of different layers possible.

The fabricated device was characterized using a modified Veeco Aurora-3 NSOM system, which uses shear-force feedback for scanning. A schematic of the measurement setup is shown in Figure 6.

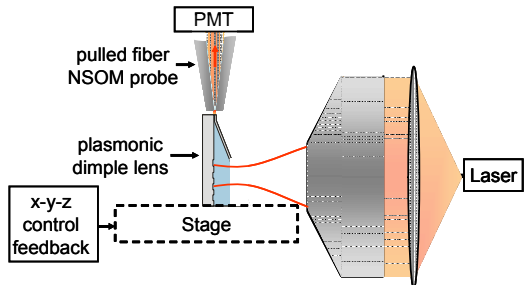

Figure 6 Schematic of the experimental setup to characterize the plasmonic dimple lens using Near-field Scanning Optical Microscopy (NSOM) technique.

The sample was mounted vertically on the NSOM stage so that the polished out-coupling edge of the device points upwards. 633 nm light from a HeNe laser was focused from the side onto the grating region of the device using a microscope objective (50X, 0.45 NA). The out-coupled light from the device was collected with a photo-multiplier tube (PMT) through a commercially available pulled-fiber NSOM probe.

The sample was scanned with the NSOM probe to acquire both the topographic and the optical data at the same time. Figures 7a,b show good correlation between the region of high intensity of light in the optical image with the dimple region of the device in the topographic scan. Furthermore the spot size formed by the dimple lens is smaller than that formed by the circular grating coupler alone, as evident from Figure 7b and 7d respectively. It is important to note that the absolute spot size measurement in Figures 7b,d is believed to be limited by the size of the pulled-fiber probe aperture (which varies between 100-150 nm) used in NSOM measurement system.

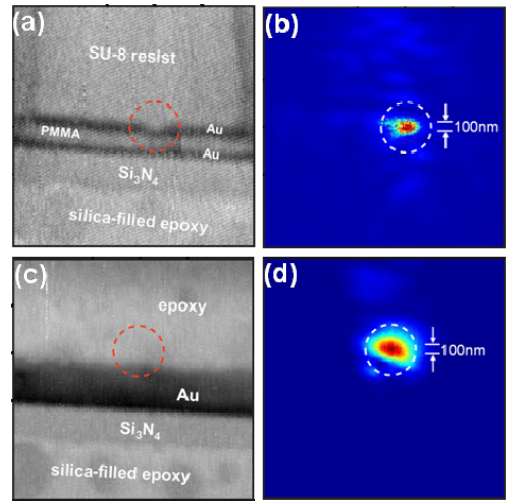

Figure 7 (a) Topographic and (b) optical images of the out-coupling facet of the plasmonic dimple lens obtained by Near-field Scanning Optical Microscopy (NSOM) technique. (c) Topographic and (d) optical images of the polishing facet of the circular grating coupler only (without a plasmonic dimple lens). All the scans are 2.5 μm x 2.5 μm.

## 4. Conclusions

In conclusion, we have designed and fabricated a novel plasmonic dimple lens structure in Au-dielectric-Au geometry for focusing visible light to the nanoscale. Comparison of measurements of the circular grating coupler alone and the circular grating coupler together with a dimple lens reveal additional focusing provided by the plasmonic dimple lens. However, the absolute measurement of the spot size produced by the dimple lens was not possible due to the aperture size limitation of the pulled-fiber NSOM probe used in the experiment. Our plasmonic lens is capable of focusing light to spot sizes smaller than the aperture size of the commercially available pulled-fiber NSOM probes.

The mechanical edge-polishing step in our fabrication sequence lends itself to be easily adopted by the process flow of the magnetic read-write head of a commercial hard drive. Thus, this structure could potentially be an excellent candidate for the Heat Assisted Magnetic Recording (HAMR) technology, where the plasmonic dimple lens would be used to focus light to provide the strong local heating required for the next generation higher density hard drives[21,22].

**Acknowledgements**

We gratefully acknowledge the support given by the NSF Nanoscale Science and Engineering Center for Scalable and Integrated Nanomanufacturing (SINAM) under the award number DMI-0327077 and the Defense Microelectronics Activity (DMEA) under the agreement number H94003-06-2-0607. We would also like to thank the UCLA Nanoelectronics Research Facility (NRF) where this fabrication process sequence is developed.